\documentclass{jfm}
\usepackage{bm}
\usepackage{graphicx}
\usepackage{epstopdf, epsfig}
\usepackage{subfigure}
\usepackage{amssymb}
\usepackage{amsmath}
\usepackage[usenames, dvipsnames]{color}
\graphicspath{{figures/}}

\shorttitle{Wall modes in magnetoconvection at high Hartmann numbers}
\shortauthor{W. Liu, D. Krasnov and J. Schumacher}

\title{Wall modes in magnetoconvection at high Hartmann numbers}

\author{Wenjun Liu\aff{1}
\corresp{\email{wenjun.liu@tu-ilmenau.de}},
            Dmitry Krasnov\aff{1}
\and    J\"org Schumacher\aff{1,2}}
\affiliation{\aff{1}Institut f\"ur Thermo- und Fluiddynamik, Technische Universit\"at Ilmenau, Postfach 100565, D-98684 Ilmenau, Germany
                \aff{2}Tandon School of Engineering, New York University, New York, NY 11201, USA}

\begin{document}

\maketitle

\begin{abstract}
Three-dimensional turbulent magnetoconvection at a Rayleigh number of $Ra=10^7$ in liquid gallium
at a Prandtl number $Pr=0.025$ is studied in a closed square cell for very strong external vertical 
magnetic fields $B_0$ in direct numerical simulations which apply the quasistatic approximation. As $B_0$ 
or equivalently the Hartmann number $Ha$ are increased, the convection flow that is highly turbulent in the 
absence of magnetic fields crosses the Chandrasekhar linear stability limit for which thermal convection is 
ceased in an infinitely extended layer and which can be assigned with a critical Hartmann number $Ha_{\rm c}$. 
Similar to rotating Rayleigh-B\'{e}nard convection, our simulations reveal subcritical sidewall modes that maintain 
a small but finite convective heat transfer for $Ha>Ha_{\rm c}$. We report a detailed analysis of the 
complex two-layer structure of these wall modes, their extension into the cell interior and a resulting sidewall boundary layer
composition that is found to scale with the Shercliff layer thickness.  
\end{abstract}


\section{Introduction}
The effect of magnetic fields on the turbulent transport of heat and momentum in turbulent convection is relevant for flow problems 
ranging from astro- and geophysics \citep{rudiger2013,Weiss2014} to numerous technological applications \citep{Davidson2016}. 
Particularly in engineering, they span a wide spectrum including materials processing,  steel casting, dendritic solidification in alloys, 
and blanket design in nuclear fusion technology.  The working fluids are then liquid molten metals with small thermal and very small 
magnetic Prandtl numbers, $Pr$ and $P_m$ which relate momentum diffusion to temperature and magnetic field diffusion, respectively. 
In addition, typically for liquid metal flows, the induced magnetic field $\bm{b}$ is assumed much smaller than the applied field $B_0$, 
which implies small magnetic Reynolds numbers $R_m \ll 1$. Therefore, the {\em quasistatic} limit of small $R_m$ is applicable with simplifications 
of the full set of magnetohydrodynamic equations \citep{Knaepen2008,Davidson2016}. It is actually also the operating point of most 
laboratory experiments on Rayleigh-B\'{e}nard convection (RBC) with external vertical \citep{Cioni2000,Burr2001,Aurnou2001} or 
horizontal magnetic fields  \citep{Fauve1981,Tasaka2016,Vogt2018}. \cite{Nakagawa1955} and \cite{Chandrasekhar1961} showed that a 
sufficiently strong vertical external magnetic field $B_0$ can suppress the onset of Rayleigh-B\'{e}nard convection. For the case of free-slip 
boundaries at the top and bottom plates, the critical Rayleigh number $Ra_{\rm c}$ of the onset of magnetoconvection in a layer, that is heated 
from below and cooled from above, is given by 
\begin{equation}
Ra_{\rm c}=\frac{\pi^2+a^2}{a^2}\left[(\pi^2+a^2)^2+\pi^2Ha^2\right]\,.
\label{Chandra}
\end{equation}
Here, $a=kH$ is the dimensionless horizontal normal mode wave number, $H$ the height of the convection layer and  $Ha$  
the dimensionless Hartmann number which is given by 
\begin{equation}
Ha=B_0 H \sqrt{\frac{\sigma}{\rho\nu}} \,.
\end{equation}
The mass density is $\rho$ and the kinematic viscosity $\nu$. Equation (\ref{Chandra}) 
is the {\em Chandrasekhar linear stability limit} of magnetoconvection and the relation holds also for no-slip boundary conditions at 
the top and bottom \citep{Chandrasekhar1961}. An equation similar to \eqref{Chandra} can be derived for a RBC layer that rotates about the vertical 
axis with a constant angular velocity $\Omega_0$ -- an alternative way to suppress the onset of convection\footnote{With the dimensionless 
Taylor number given by $Ta=4\Omega_0^2 H^4/\nu^2$, equation (\ref{Chandra}) for rotating convection follows by the substitution 
$(\pi^2+a^2)Ha^2=Ta$.}. It is well-known since the experiments in water or oil of \cite{Zhong1991}, \cite{Ecke1992}, \cite{Liu1999}, and \cite{King2012}
and the linear stability analyses of \cite{Goldstein1993,Goldstein1994} that the existence of sidewalls in closed and rotating cylindrical cells can 
destabilize convection. For recent direct numerical simulations of rotating liquid metal convection flows, we refer additionally to \cite{Horn2017}. 
In other words, convection is present for $Ra<Ra_{\rm c}$ in form of subcritical modes attached to the sidewalls which are denoted as {\em wall modes} in the following.  
This result together with the close analogy to rotating RBC sets the motivation for the present study.
    
We investigate the impact of a strong vertical magnetic field on a liquid metal convection flow in a closed square cell of
aspect ratio 4 by a series of three-dimensional DNS. Therefore liquid metal convection at 
fixed Rayleigh and Prandtl numbers, $Ra=10^7$ and $Pr=0.025$, (the flow is highly turbulent in absence of a 
magnetic field) is driven to cross the Chandrasekhar stability limit (\ref{Chandra}) by a stepwise increase of $B_0$. The 
linear stability limit is reached at $Ha_{\rm c}\approx\sqrt{Ra}/\pi\approx 1000$ for the chosen $Ra$ and $Pr$. Our numerical studies 
show that convective heat transfer is still present for Hartmann numbers up to $Ha=2 Ha_{\rm c}$. We also demonstrate by a scale-refined 
analysis in concentric subvolumes that the transport of heat 
and momentum is maintained by subcritical flow modes which are attached to the sidewalls, similar to rotating RBC. Furthermore,  
the spatial organization and structure of these (quasisteady) wall modes is studied in detail. The modes are found here to take the form of circulation rolls that are 
increasingly closer attached to the sidewalls as $Ha$ grows and form a two-layer boundary flow that scales with the Shercliff thickness, 
a characteristic boundary scale for a shear flow in a transverse magnetic field \citep{Shercliff1953}. Wall modes in
magnetoconvection were investigated in linear stability analysis by \cite{Houchens2002} in cylindrical cells and predicted by an asymptotic 
theory along a single straight vertical sidewall between free-slip boundaries at the top and bottom by \cite{Busse2008}.     
Their existence and their complex spatial structure is analysed here for the first time in fully resolved direct numerical simulations 
of a typical laboratory experiment configuration that starts with fully developed convective turbulence at high Rayleigh number.

\section{Numerical model}\label{sec:method}
We solve the three-dimensional equations of magnetoconvection in a closed square box,
with $x,y$ as horizontal and $z$ as vertical directions, using the quasistatic limit \citep{Zuerner2016}. 
They couple the velocity field ${\bm u}$ with the temperature field $T$ and the external magnetic field ${\bm B}=B_0{\bm e}_z$.
The equations are made dimensionless by using height of the cell $H$, the free-fall velocity $U_{\rm f}=\sqrt{g \alpha \Delta T H}$,
the external magnetic field strength $B_0$, and the imposed temperature difference $\Delta T=T_{\rm bottom}-T_{\rm top}$. Four dimensionless
control parameters are contained: the Rayleigh number $Ra$, the Prandtl number $Pr$, the Hartmann number $Ha$ and the aspect ratio
$\Gamma = L/H$ with $L_x = L_y = L$. The equations are given by
\begin{align}
\label{ceq}
{\bm \nabla}\cdot {\bm u}&=0\,,\\
\label{nseq}
\frac{\partial  \bm u}{\partial  t}+(\bm u \cdot \bm \nabla){\bm u}
&=-\bm\nabla p+\sqrt{\frac{Pr}{Ra}} \left[\bm\nabla^2 \bm u+Ha^2 (\bm j\times \bm e_z)\right] + T \bm e_z\,,\\
\frac{\partial  T}{\partial  t}+(\bm u \cdot \bm \nabla) T
&=\frac{1}{\sqrt{Ra Pr}} \bm\nabla^2 T\,,
\label{pseq}
\end{align}
where $p$ is the pressure field. The Rayleigh number is $Ra=g\alpha\Delta T H^3/(\nu\kappa)$ and the Prandtl number $Pr=\nu/\kappa$.
The variable $g$ stands for the  acceleration due to gravity, $\alpha$ is the thermal expansion coefficient,
$\kappa$ the thermal diffusivity, and $\rho$ the mass density. No-slip boundary conditions for the velocity 
are applied at all walls. The sidewalls are thermally insulated and the top and bottom plates are held fixed at
$T=-0.5$ and 0.5, respectively. All walls are in addition perfectly electrically insulating such that the electric
current density ${\bm j}$ has to form closed field lines inside the cell. Together with the Ohm law,
${\bm j}=-\bm \nabla \phi + (\bm u\times \bm e_z)$ and divergence-free currents ${\bm \nabla}\cdot {\bm j}=0$,
this results in an additional Poisson equation for the electric potential $\phi$ which is given by
\begin{equation}
\bm \nabla^2 \phi=\bm \nabla \cdot (\bm u\times \bm e_z) \text{  with  } {\partial \phi}/{\partial n}=0\, \text{ on the walls}.
\end{equation}
This completes the quasistatic magnetoconvection model. The equations are solved by a second-order finite difference
method on a non-uniform Cartesian mesh, discussed in detail in \cite{Krasnov2011}. Table \ref{tab:kd} summarizes
the most important parameters of our DNS runs and reports the (turbulent) momentum transfer quantified by the Reynolds number
$Re=u_{\rm rms} \sqrt{Ra/Pr}$ with $u_{\rm rms}=\langle u_x^2+u_y^2+u_z^2\rangle_{V,t}^{1/2}$ and (turbulent) heat transfer measured
by the Nusselt number $Nu=1+\sqrt{RaPr}\langle u_zT\rangle_{V,t}$ with $\langle\cdot\rangle_{V,t}$ being a combined volume
and time average \citep{Scheel2016}. The values of $Re$ and $Nu$ of run 1 are comparable with those from \cite{Scheel2016}
for RB convection in mercury at $Pr=0.021$ in a closed cylindrical cell at $\Gamma=1$ where
$Nu=10.11 \pm 0.05$ and $Re=8450 \pm 100$ is reported.

Beside the thermal boundary layer (BL) thickness $\delta_T=1/(2 Nu)$ and the viscous boundary layer thickness
$\delta_v=1/(4\sqrt{Re})$, two further BL thicknesses are relevant for the problem at hand. These are the Hartmann
layer thickness $\delta_{\text{Ha}} = a_1/Ha$ \citep{Hartmann1937} at the top and bottom plates and the Shercliff
layer thickness $\delta_{\text{Sh}} = a_2/\sqrt{Ha}$ \citep{Shercliff1953} at the sidewalls. Constants $a_1$ and $a_2$ are of order
${\cal O}(1)$ and geometry-dependent. For simplicity, we set $a_1=a_2=1$ for the following. The grid resolution inside
the Hartmann and viscous boundary layers is also listed in table 1 by $N_{BL}$.   

\begin{table}
\small{
\begin{center}
\def~{\hphantom{0}}
\begin{tabular}{cccccccc}
Run  &  $Ha$ & $Ra/Ra_{\rm c}$ & $N_x\times N_y \times N_z$ &  $Nu$  &   $Re$  &  $N_{BL}$ &  Runtime  \\[3pt]
1  &     0   &  $\infty$  &  $\;2048\times 2048\times 512\;$ &  $\;9.75\pm 0.05\;$  &  $\;7946\pm 19\;$  &  18  &   31\\                      
2  &   200   &  25.33     &  $\;\;2048\times 2048\times 512\;\;$ &  $\;\;7.69\pm 0.12\;\;$  &  $\;\;3532\pm 27\;\;$   &  29   &  31\\
3  &   500   &   4.05     &  $\;\;2048\times 2048\times 512\;\;$ &  $\;\;4.11\pm 0.05\;\;$  &  $\;\;1714\pm 15\;\;$   &  14  &   31\\
4  &  1000   &   1.01     &  $\;2048\times 2048\times 512\;$ &  $\;1.41$  &  $\;565\pm 1\;$    &  8   &   31\\
5  &  1500   &   0.45     &  $\;2048\times 2048\times 512\;$ &  $\;1.28$  &  $\;425$    &  8   &   31\\
6  &  2000   &   0.25     &  $\;2560\times 2560\times 640\;$ &  $\;1.15$  &  $\;287$    &  8   &   31\\
\end{tabular}  
\caption{Parameters of the simulations. The Prandtl number is fixed to $Pr=0.025$, the Rayleigh number to $Ra=10^7$,
and the aspect ratio to $\Gamma=4$. The Hartmann number $Ha$, the ratio $Ra/Ra_{\rm c}$ with $Ra_{\rm c}\approx \pi^2Ha^2$
(ratios $Ra/Ra_{\rm c}<1$ imply convection beyond the Chandrasekhar limit), the grid resolution, the Nusselt number $Nu$,
the Reynolds number $Re$ are given. We also list the number of horizontal grid planes inside the Hartmann layer with
thickness $\delta_{\text{Ha}}$. In case of $Ha=0$, we substitute the Hartmann layer thickness by the viscous boundary 
layer thickness $\delta_v$. Finally, the total runtime is given in free fall time units, $T_{\rm f}=H/U_{\rm f}$.}
\label{tab:kd}
\end{center}
} 
\end{table}

\section{Results}
\begin{figure}
\centering
\includegraphics[height=9cm]{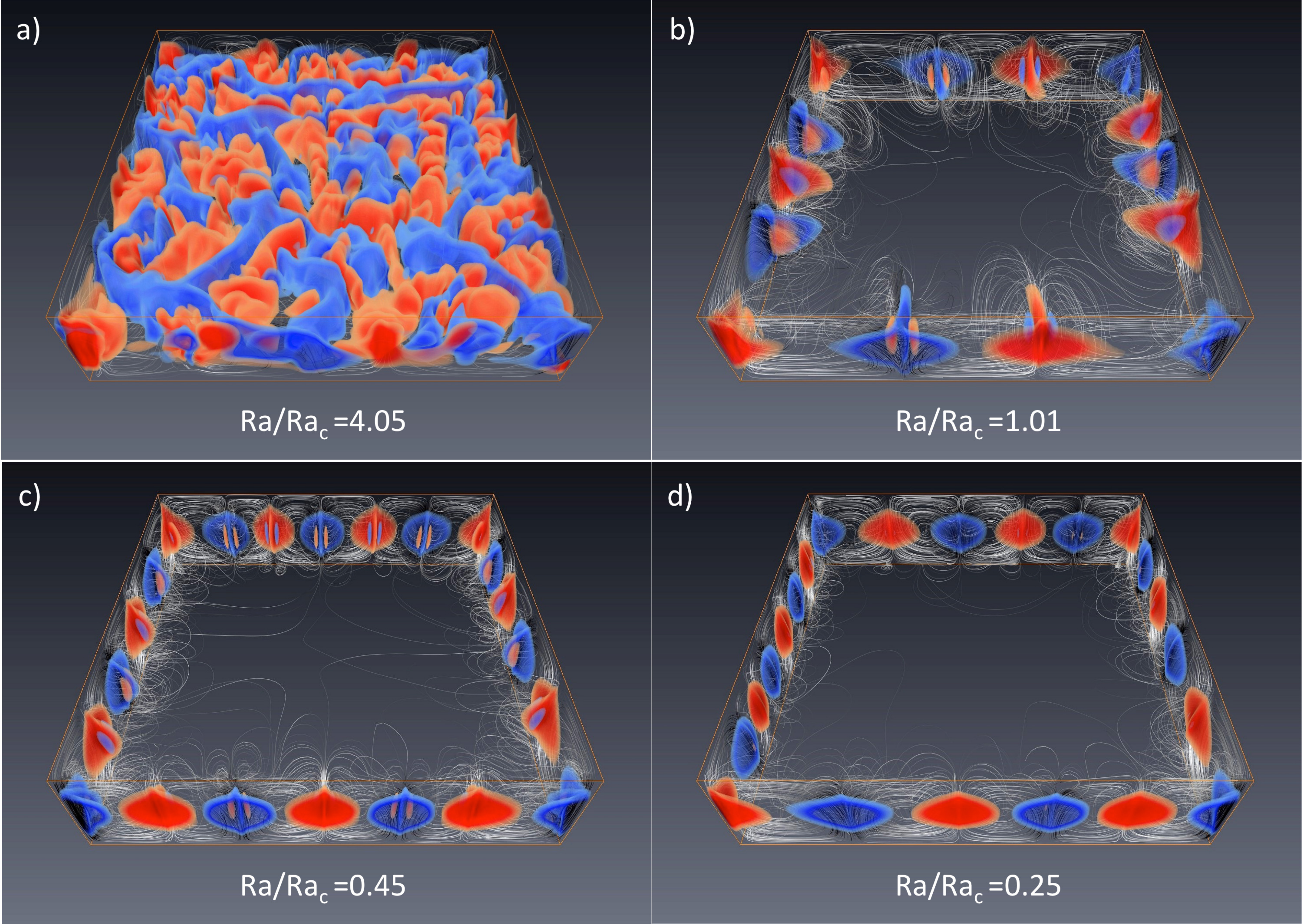}
\caption{Snapshots of the magnetoconvection flow at (a) $Ha=500$. (b) $Ha=1000$. (c) $Ha=1500$. (d) $Ha=2000$. 
We display isosurfaces of $u_z=\pm 0.01$ together with field lines of the velocity field that highlight the circulation rolls 
at the sidewalls. The ratio $Ra/Ra_{\rm c}$ is indicated in each of the panels with $Ra_{\rm c}=\pi^2 Ha^2$. For ratios of $Ra/Ra_{\rm c}<1$ 
the linear stability theory predicts a purely diffusive transport in a fluid layer with ${\bm u}=0$.}
\label{fig:1}
\end{figure}

\subsection{Spatial structure of sidewall modes} 
Figure \ref{fig:1} displays snapshots of the velocity field structure of the magnetoconvection flows. Isosurfaces of
the vertical velocity component and field lines of the velocity field are shown for runs at $Ha\ge 500$. For small
external magnetic field strength, a cellular structure of up- and downwelling flows is observed that fills the entire cell.
A sufficiently strong external magnetic field that corresponds here to $Ha \ge 1000$ expels convective motion from
the interior of the cell where heat is then transported solely by diffusion. We note that the critical Hartmann number,
which corresponds to the Chandrasekhar linear stability limit, is given by $Ha_{\rm c} = \sqrt{Ra}/\pi \approx 1007$ for the present
Rayleigh number of $Ra = 10^7$. Figures \ref{fig:1}(b)--(d) display the structure of the wall modes, which consist of alternating
up- and downwelling flow regions attached to the sidewalls. They correspond to neighboring circulation rolls which
do not move along the sidewalls or oscillate as in rotating convection for the total integration times that we could run the simulations. 
The wall modes are ever closer attached to the sidewalls
as $Ha$ grows from $1000$ to $2000$. Interestingly, vertical velocity maxima reach out into the bulk in form of tongue-like
filaments which might be a relic from the turbulent flow pattern, a point that will be analysed more closely in subsection 3.3.
\begin{figure}
\centering
\setlength{\unitlength}{1cm}
\begin{picture}(5.0,3.5)
\put(0,0){\includegraphics[height=3.5cm]{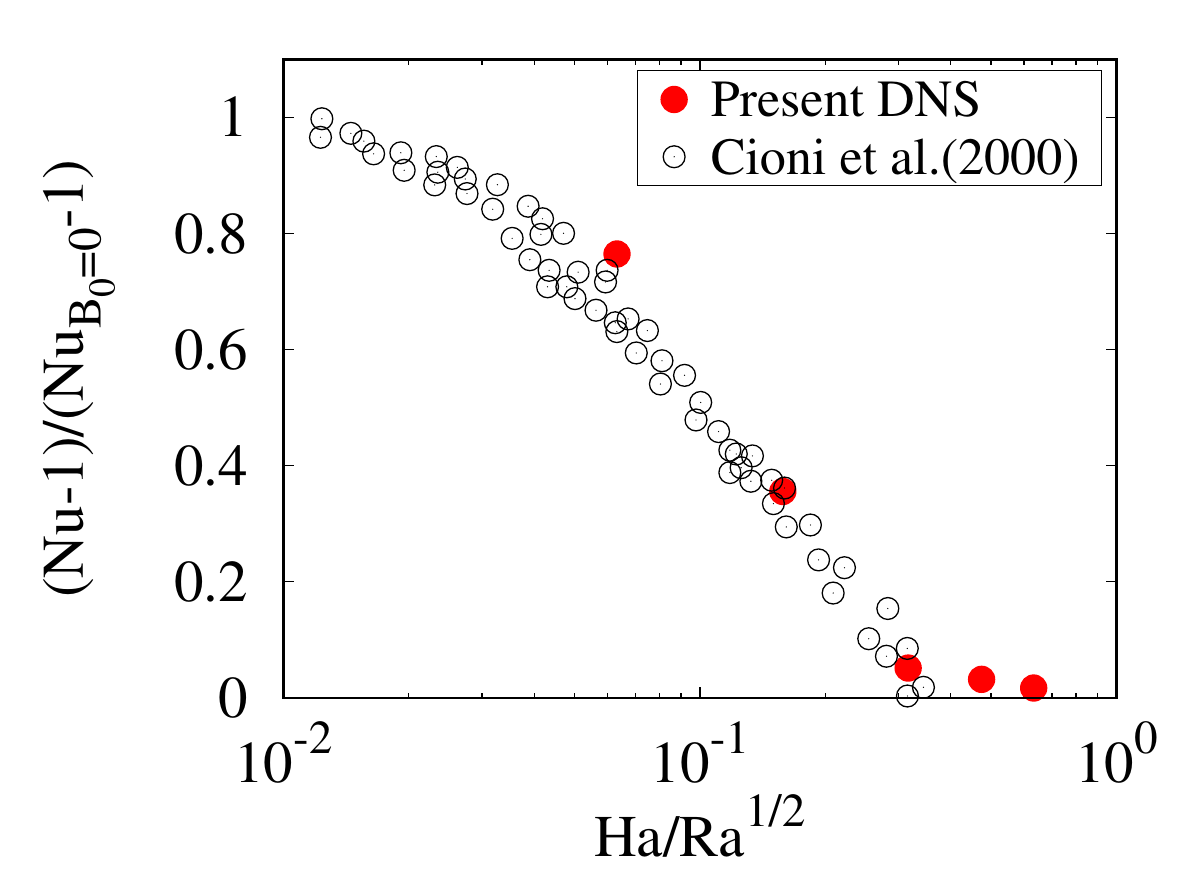}}
\put(0,3.1){\textit{a})}
\end{picture}
\begin{picture}(5.0,3.5)
\put(0,0){\includegraphics[height=3.5cm]{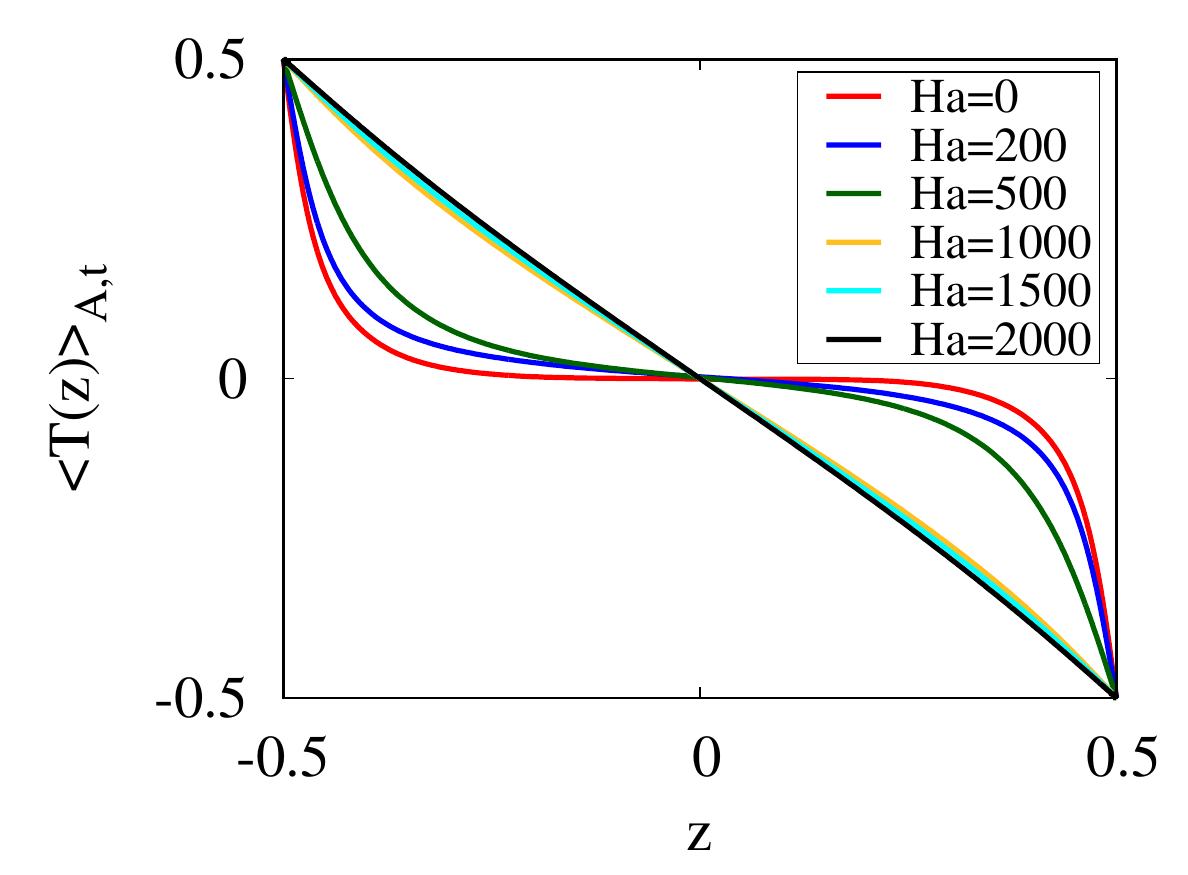}}
\put(0,3.1){\textit{b})}
\end{picture}
\\
\begin{picture}(5.0,3.5)
\put(0,0){\includegraphics[height=3.5cm]{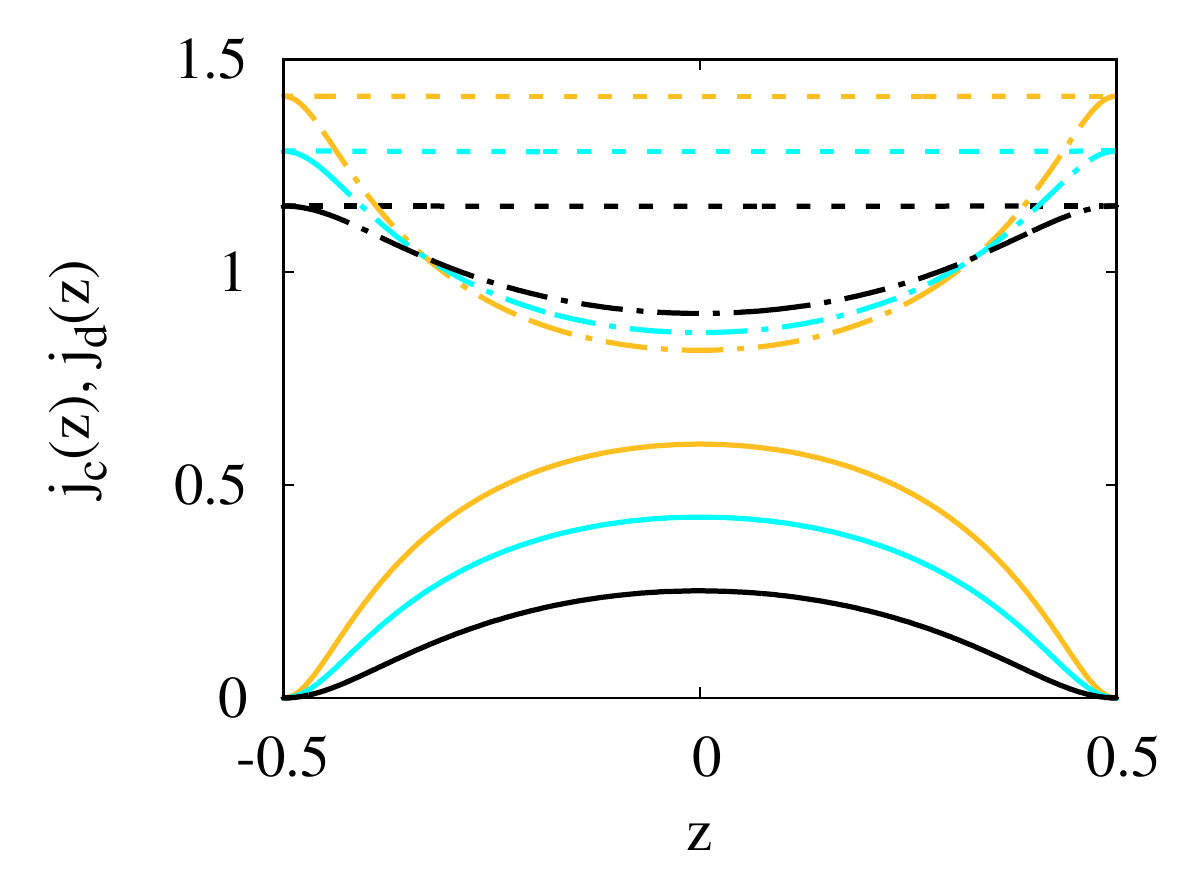}}
\put(0,3.1){\textit{c})}
\end{picture}
\begin{picture}(5.0,3.5)
\put(0,0){\includegraphics[height=3.5cm]{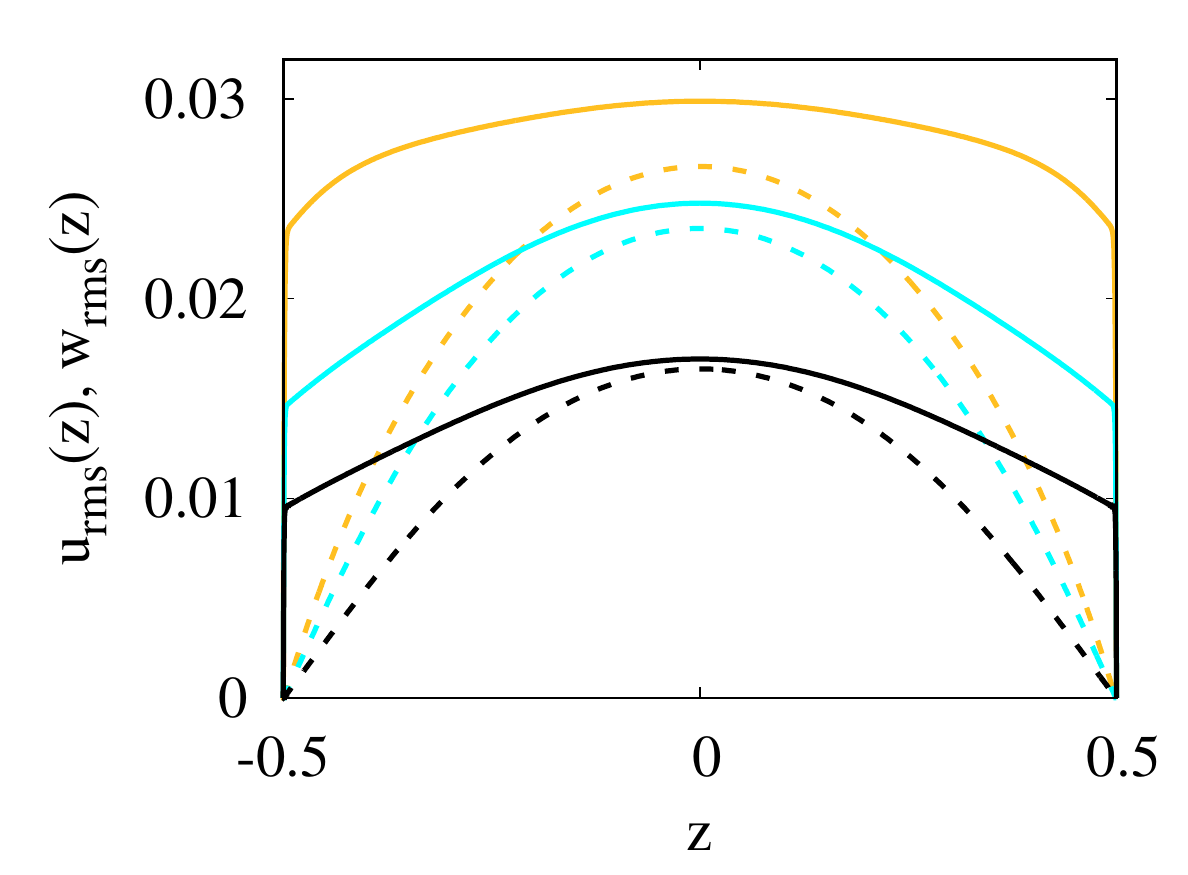}}
\put(0,3.1){\textit{d})}
\end{picture}
\caption{Global transport of heat and momentum. (a) Nusselt number $Nu$ normalized with respect to the Nusselt number 
value at $B_0=0$ versus $Ha/\sqrt{Ra}$. Present data are given by red filled circles.
A part of the data of the experiment by \cite{Cioni2000} is added (open circles). (b) Mean profiles of temperature $T(z)=\langle T(z)\rangle_{A,t}$. 
(c) Mean profiles of convective heat current, $j_{\rm c}(z)$ (solid lines), and diffusive heat current, $j_{\rm d}(z)$ (dash-dotted lines) 
for various Hartmann numbers. Both currents are defined in \eqref{Nusselt}. The dotted lines in (c) correspond to 
$Nu(z)=j_{\rm c}(z)+j_{\rm d}(z)=const.$ (d) Mean profile of the root-mean-square (rms) for the full velocity field (solid line) 
and the vertical velocity component (dashed line).}
\label{fig:z_profile}
\end{figure}
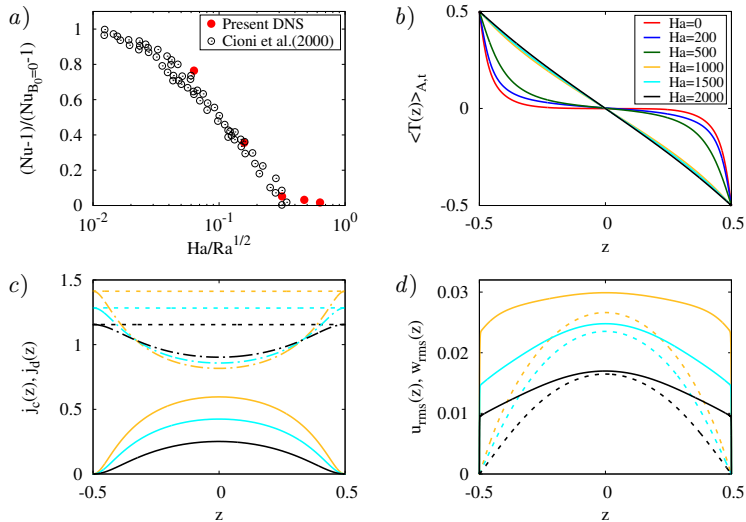

\subsection{Global and near-sidewall transport of heat and momentum}   
Table \ref{tab:kd} shows that the Nusselt number is still larger unity for runs 4--6 beyond the Chandrasekhar limit.
With growing $Ha$ the global transport converges to the pure diffusion case at  $Nu=1$. The strong temporal velocity fluctuations of the 
highly inertial low--$Pr$ turbulence for absent or weak magnetic fields decrease then to vanishingly small values and result in a practically 
quasisteady laminar wall mode flow. The global analysis of transport is refined in figure~\ref{fig:z_profile} by vertical mean profiles 
taken with respect to horizontal plane $A=\Gamma^2$ and time, $\langle\cdot\rangle_{A,t}$. Figure~\ref{fig:z_profile}(a) compares our 
Nusselt numbers directly to the experiments in a closed cylinder by \cite{Cioni2000}. It can be seen that their data record coincides well 
the DNS below the Chandrasekhar limit. Deviations are observed for the highest magnetic fields where the data points of the 
simulation branch off due to the existence of wall modes. Figure~\ref{fig:z_profile}(b) shows the temperature profiles for all six data sets. 
The well-mixed bulk region for the turbulent cases at $Ha < 1000$ changes to an almost linear diffusion-dominated profile for $Ha\ge 1000$. 
The Nusselt number is decomposed into two terms that stand for the convective and diffusive heat fluxes across the layer,
\begin{equation}
Nu(z)=j_{\rm c}(z)+j_{\rm d}(z)=\sqrt{Ra Pr}\, \langle u_z T\rangle_{A,t}-\frac{\partial \langle T\rangle_{A,t}}{\partial z}\,.
\label{Nusselt}
\end{equation}
Figure~\ref{fig:z_profile}(c) displays the profiles of $j_{\rm c}(z)$ and $j_{\rm d}(z)$ for the three runs with $Ha\ge 1000$ as well as their sum 
(which has to be constant and equal to $Nu$). It is seen that $j_{\rm d}(z) >j_{\rm c}(z)$. The increasing suppression of fluid turbulence is also 
demonstrated in figure~\ref{fig:z_profile}(d) by the root-mean-square (rms) velocity profiles which are given for the {\em quasisteady} cases 
for $Ha\ge 1000$ by $u_{\rm rms}(z)=\langle u_x^2+u_y^2+u_z^2\rangle_{A}^{1/2}$ and $w_{\rm rms}(z) =\langle u_z^2\rangle_{A}^{1/2}$. 
It is not only that total fluctuation level decreases then as a whole, but also that the vertical velocity fluctuations provide an increasing fraction 
to the total fluctuation magnitude for $Ha\ge Ha_{\rm c}$. Furthermore, we find that the ratio of the root mean square values taken in the full 
cell volume, $w_{\rm rms}/u_{\rm rms}= 0.55, 0.78, 0.90, 0.71, 0.82$, and 0.83 for the six simulation runs. 

\begin{figure}
\centering
\setlength{\unitlength}{1cm}
\begin{picture}(4.4,3.3)
\put(0,0){\includegraphics[trim={0.7cm 0cm 0cm 0cm},clip,width=4.5cm]{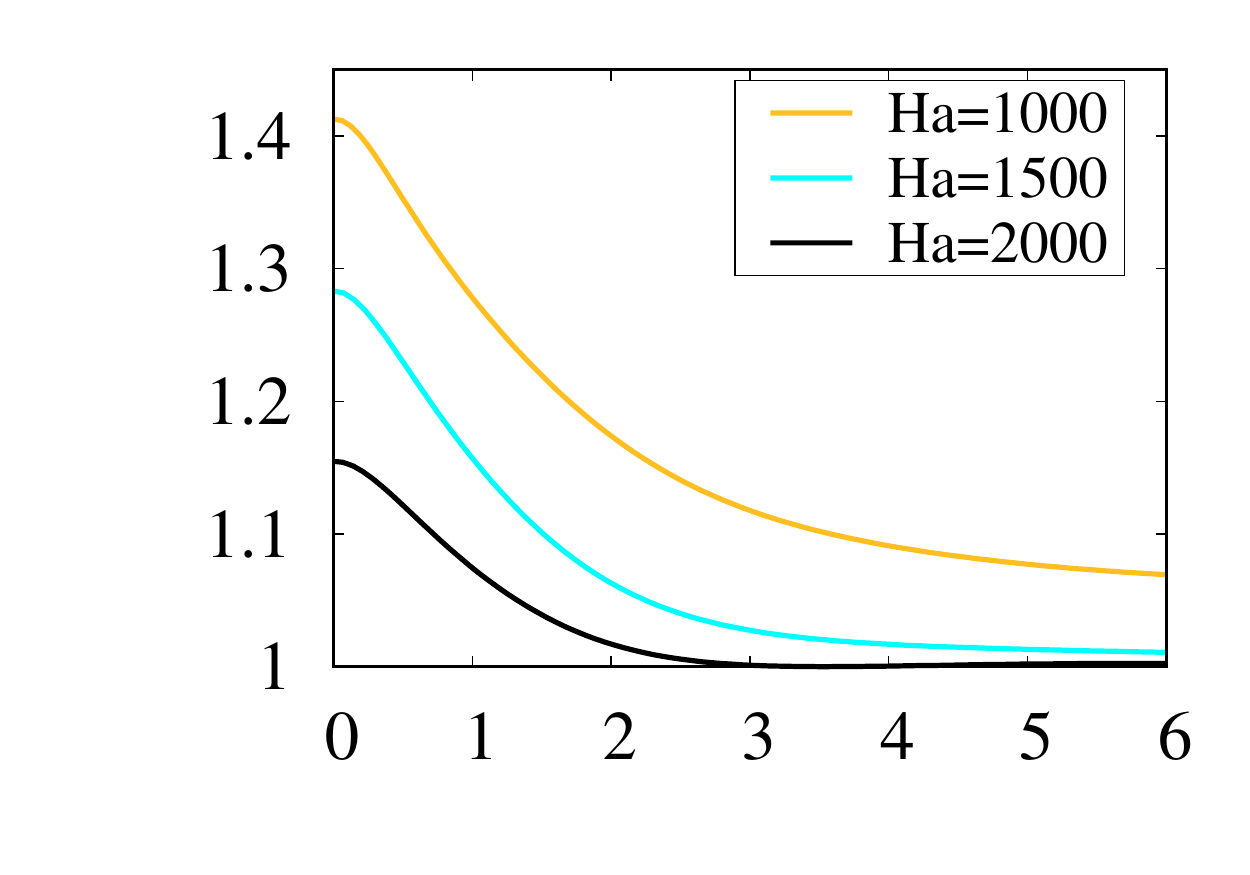}}
\put(2.2,0.05){$r/\delta_{\text{Sh}}$}
\put(0.1,1.3){\rotatebox{90}{$\langle Nu \rangle_S$}}
\put(0.05,3){\textit{a})}
\end{picture}
\begin{picture}(4.4,3.3)
\put(0,0){\includegraphics[trim={0.7cm 0cm 0cm 0cm},clip,width=4.5cm]{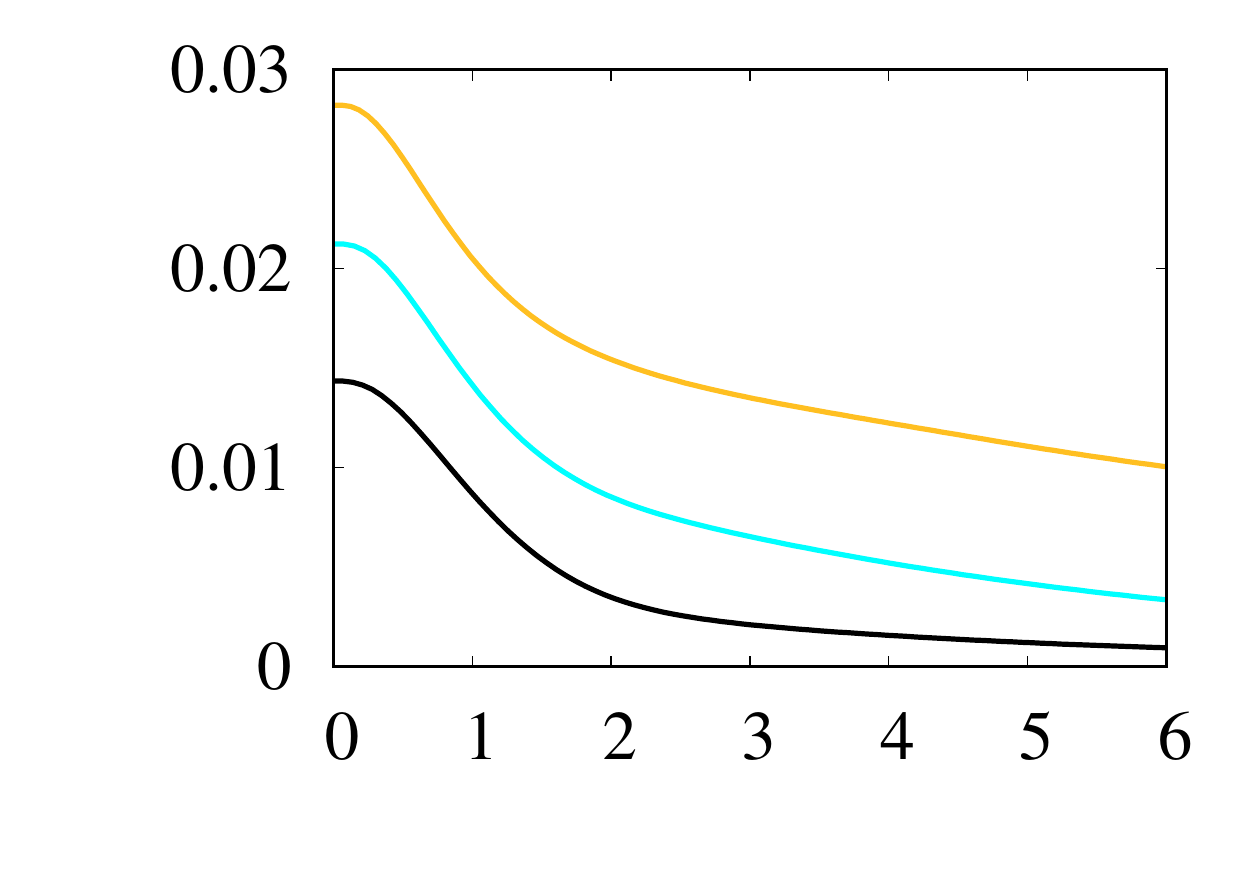}}
\put(2.15,0.05){$r/\delta_{\text{Sh}}$}
\put(-0.1,1.25){\rotatebox{90}{$\sqrt{\langle u_i^2\rangle_S}$}}
\put(-0.1,3){\textit{b})}
\end{picture}
\begin{picture}(4.4,3.3)
\put(0,0){\includegraphics[trim={0.7cm 0cm 0cm 0cm},clip,width=4.5cm]{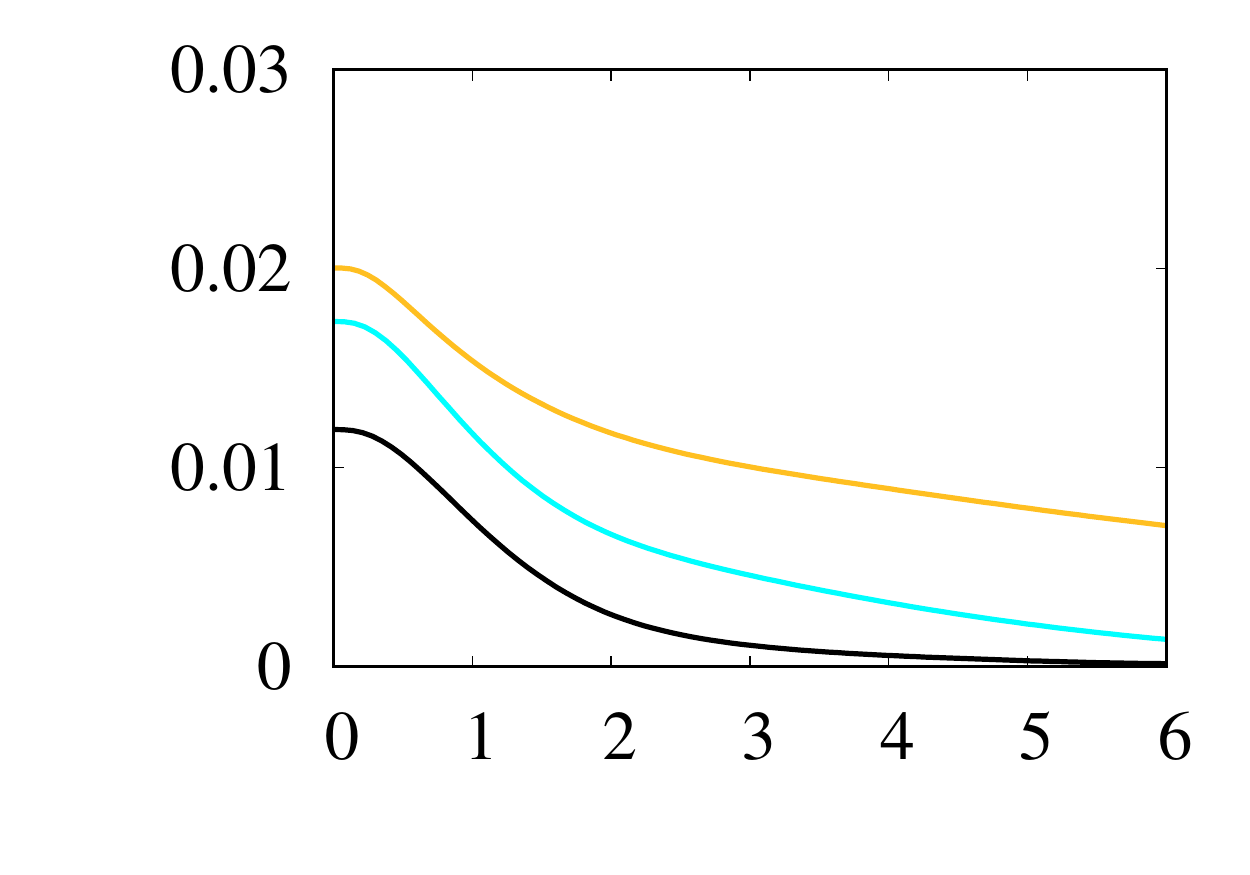}}
\put(2.15,0.05){$r/\delta_{\text{Sh}}$}
\put(-0.1,1.25){\rotatebox{90}{$\sqrt{\langle u_z^2\rangle_S}$}}
\put(-0.1,3){\textit{c})}
\end{picture}
\caption{Near-sidewall transport of heat and momentum beyond the Chandrasekhar stability limit. Nusselt number and root mean 
square velocities are determined over successively smaller horizontal cross section $S$. Profiles are plotted versus a sidewall-normal 
distance $r$ in units of the Shercliff layer thickness $\delta_{\text{Sh}}$. (a) Nusselt number $\langle Nu\rangle_S$. (b) Root mean 
square velocity $\langle u_i^2\rangle_{S}^{1/2}$=$\langle u_x^2+u_y^2+u_z^2\rangle_{S}^{1/2}$. (c) Root mean square vertical 
velocity component $\langle u_z^2\rangle_{S}^{1/2}$. Data are $Ha=1000, 1500$ and 2000 as indicated in the legend in (a).}
\label{fig:3}
\end{figure}
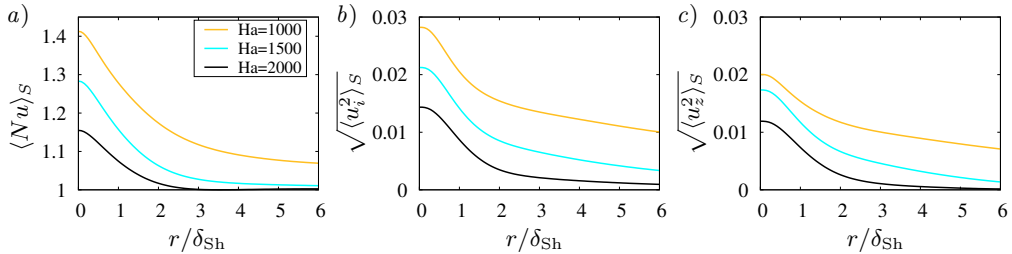

The importance of the wall modes for the transport of heat and momentum beyond the Chandrasekhar limit is highlighted in figure \ref{fig:3}
where we determine Nusselt number and root mean square velocities over successively smaller concentric cross section areas 
$S=[r, \Gamma-r]\times [r,\Gamma-r]$ with $r$ being the sidewall-normal distance. These averages are indicated by $\langle\cdot\rangle_S$. 
As visible in figure 3, the transport decreases significantly within a few Shercliff layer thicknesses $\delta_{\text{Sh}}=1/\sqrt{Ha}$ thus further 
supports the hypothesis that the transport for $Ha>Ha_{\rm c}$ is connected to the wall modes. It is also seen that the heat transfer drops to 
the diffusive lower bound of $Nu=1$ for $r\gtrsim 2\delta_{\text{Sh}}$ for the highest Hartmann number. All velocity profiles drop significantly 
towards the center of the cell.

\subsection{Wall-mode structure together with thermal and Shercliff boundary layers}
We proceed with an analysis of the viscous and thermal boundary layers in conjunction with the wall-modes. 
In table \ref{Tab2}, we list all important BL thicknesses. The thermal boundary layer thickness $\delta_T=1/(2 Nu)$
at the top and bottom approaches 0.5 in agreement with $Nu\to 1$ as $Ha$ grows. The mean thermal BL thickness 
at the sidewall, $\delta_T^{\rm (sw)}$ is obtained from profiles of the temperature fluctuations, $\theta({\bm x},t)=T({\bm x},t)
-\langle T(z)\rangle_{A,t}$, with respect to the sidewall-normal coordinate $r$. These profiles are calculated for $Ha\ge 1000$ 
only, i.e., when the dominantly vertical up- and downflows are attached to the sidewalls. They are always obtained as an 
average over all four sidewalls. The value of $\delta_T^{\rm (sw)}$ is determined by a standard slope method, 
i.e., as the intersection point of the horizontal line drawn through the mean value of $\langle \theta(r)\rangle_{\rm sw}$ in 
the bulk and a tangent which is fitted to the same profile very close to the sidewall. The corresponding values decrease as 
$Ha$ grows and are given in table \ref{Tab2}.
   
In the presence of a strong $B_0$, the standard viscous boundary layer thickness $\delta_v$ has to be substituted
by the Hartmann layer thickness $\delta_{\text{Ha}} = 1/Ha$ at the top and bottom. As seen in table \ref{Tab2},
the Hartmann layers become extremely thin and their appropriate resolution makes these DNS very demanding.
The viscous sidewall layers are also affected by the external magnetic field. Here the fluid motion in horizontal ($x$,$y$)-directions, 
i.e. transverse to the sidewall-parallel magnetic field, is affected and Shercliff layers with thickness $\delta_{\rm Sh}$ are formed. The Shercliff
thickness will be chosen as the length-scale in which the wall modes are measured. The latter ones establish
a complex flow structure at the sidewalls which is quantified in figure \ref{fig:ha2000} for the highest external field at $Ha=2000$.
We observe the alternating up- and downflows of warmer and colder fluid, respectively (see figure~\ref{fig:ha2000}(a,b)).
The tongue-like structure consists of three thin counter-flowing jets (up-down-upwelling or down-up-downwelling)
which arise due to the incompressibility condition (see figures \ref{fig:ha2000}(b) and \ref{fig:1}). The velocity
amplitude inside the modes is still remarkably large with a maximum of $u_z \sim 0.1$ as seen in figure \ref{fig:ha2000}(d).

Figures~\ref{fig:ha2000}(b,c) shows a two-layer structure of the wall modes on the basis of the vertical velocity 
component and heat transport. Figure \ref{crossover} supports this observation further by an analysis of the convective heat transfer
$\langle Nu\rangle_S-1=\sqrt{Ra Pr} \langle u_z T\rangle_S$ (see also figure~3(a)). For all runs at $Ha \ge 1000$,
two exponential decays laws can be observed. These decays separate two different layers: the bulk region dominated by 
diffusive heat transport and the inner near-wall region with residual convective flow motion, particularly well observable 
at the highest $Ha=2000$. We have verified that the pronounced minimum at this largest Hartmann number persists 
for finer computational grids by shorter test reruns at higher resolutions.
The crossover distance $r_{\rm cr}$, identified as the intersection point of both exponential fits, 
is found at a fixed ratio to the Shercliff layer thickness $\delta_{\text{Sh}}$ for the three runs, as given in the table \ref{Tab2}. 
This result implies that the inner section of the sidewall layer is on average of Shercliff-type despite the alternating pattern of 
horizontal up- and downflows. The distance $r_{\rm cr}$ matches the point where the thin tongue-like vertical flows in 
figure \ref{fig:ha2000}(b,c) appear. Interestingly, the exponential fit $Nu-1 \approx A \times \exp(-\beta r)$ of the inner sublayer for 
$r \le r_{\rm cr}$ results in spatial decay rates $\beta$ being a fixed ratio to the interaction parameter (or Stuart number) 
$N = Ha^2/\sqrt{Ra/Pr}$. This parameter relates Lorentz to inertial forces and, in the present DNS series, underlines the dominance 
of Lorentz forces at $Ha \ge 1000$.
\begin{table}
\small{
\begin{center}
\def~{\hphantom{0}}
\begin{tabular}{ccccccccc}	
Run  & $Ha$ & $\delta_T$ & $\delta^{\rm (sw)}_T$ & $\delta_{\rm Ha}$ & $\delta_{\rm Sh}$ & $\delta^{\rm (sw)}_T/\delta_{\rm Sh}$ & $r_{\rm cr}/\delta_{\rm Sh}$ &$\beta/N$ \\[3pt]
1   & 0        & 0.051 &   --       & $\infty$  & $\infty$       & --        & --       & --\\    
2   & 200    & 0.065 &   --       & 0.005     & $0.071$     & --        & --       & --\\
3   & 500    & 0.122 &   --       & 0.002     & $0.045$     & --        & --       & --\\
4   & 1000   & 0.355 & 0.338  & 0.0010   & $0.032$    & 10.69  & 3.46  & 0.30\\
5   & 1500   & 0.391 & 0.260  & 0.0007   & $0.026$    & 10.07  & 3.46  & 0.30\\
6   & 2000   & 0.435 & 0.227  & 0.0005   & $0.022$    & 10.15  & 3.47  & 0.30\\
\end{tabular}  
\caption{Summarizing list of different boundary layer thicknesses which can be obtained in the magnetoconvection flow in a closed cell. We list thermal BL
thicknesses at top/bottom and sidewalls as well as Hartmann and Shercliff layer thicknesses. For the runs with wall modes we also list the ratios 
$r_{\rm cr}/\delta_{\rm Sh}$ and $\beta/N$ to quantify a two-layer structure. Here, $r_{\rm cr}$ is the crossover width of the two spatial decays laws of convective 
heat flux of the wall modes. The exponential decay of the inner layer is measured by exponent $\beta$ which is found in a fixed ratio to the corresponding 
interaction parameter (or Stuart number) $N=Ha^2/\sqrt{Ra/Pr}$.}
\label{Tab2}
\end{center}
} 
\end{table}
\vskip-5mm

\begin{figure}
\centering
\setlength{\unitlength}{1cm}
\begin{picture}(6.6,2.5)
\put(0,0){\includegraphics[trim={0cm 5.5cm 0cm 7cm},clip,width=6.7cm]{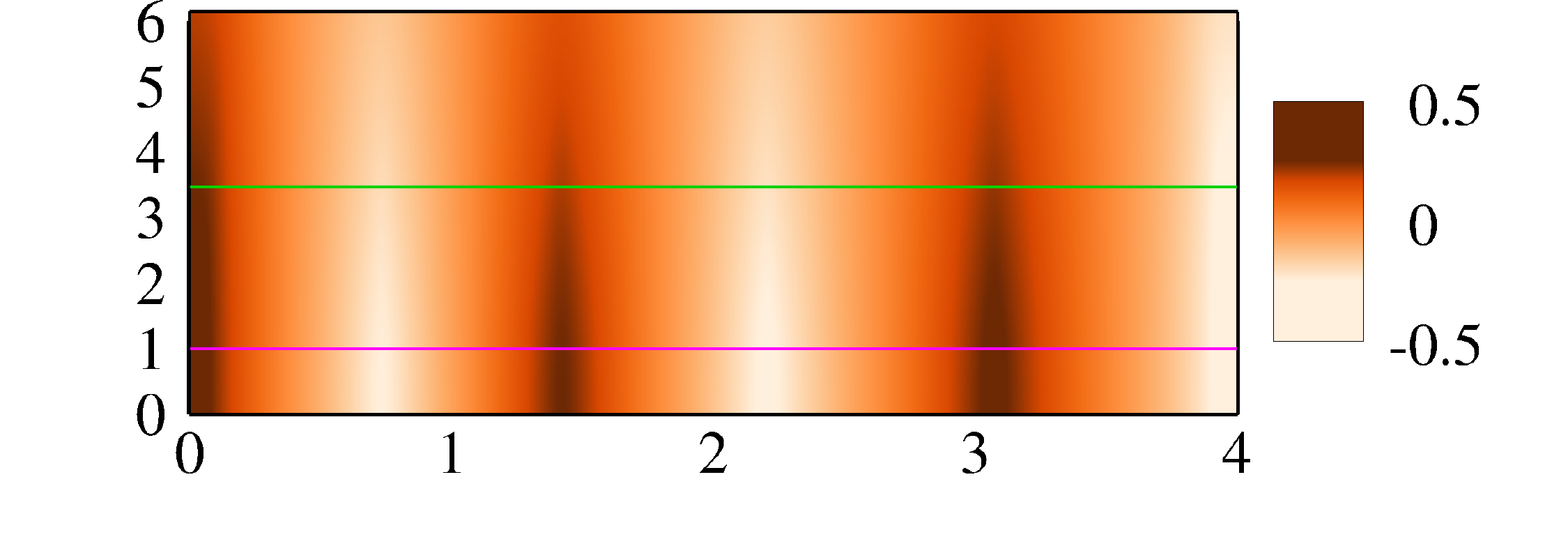}}
\put(3.0,0){$x$}
\put(0.14,1){\rotatebox{90}{$y/\delta_{\rm Sh}$}}
\put(0,2.1){\textit{a})}
\end{picture}
\hspace{0.1cm}
\begin{picture}(6.5,2.5)
\put(0,0){\includegraphics[trim={0cm 5.5cm 0cm 7cm},clip,width=6.7cm]{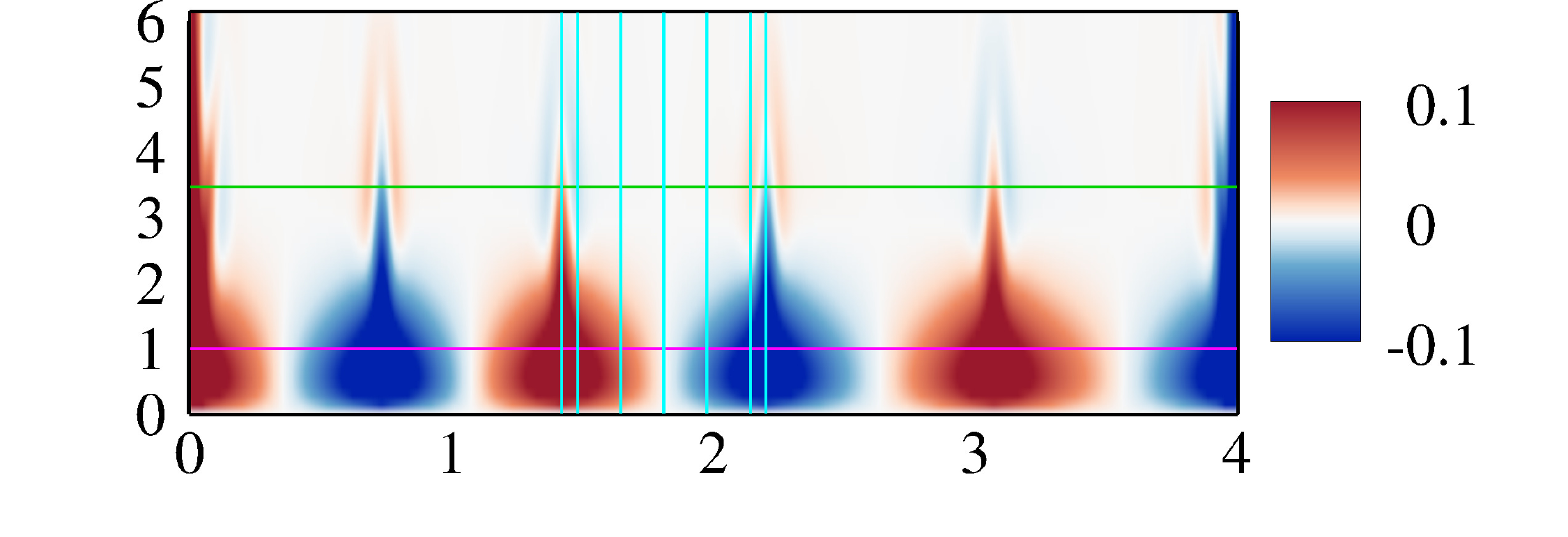}}
\put(3.0,0){$x$}
\put(0.14,1){\rotatebox{90}{$y/\delta_{\rm Sh}$}}
\put(0,2.1){\textit{b})}
\end{picture}
\\
\begin{picture}(6.5,2.5)
\put(0,0){\includegraphics[trim={0cm 5.5cm 0cm 7cm},clip,width=6.7cm]{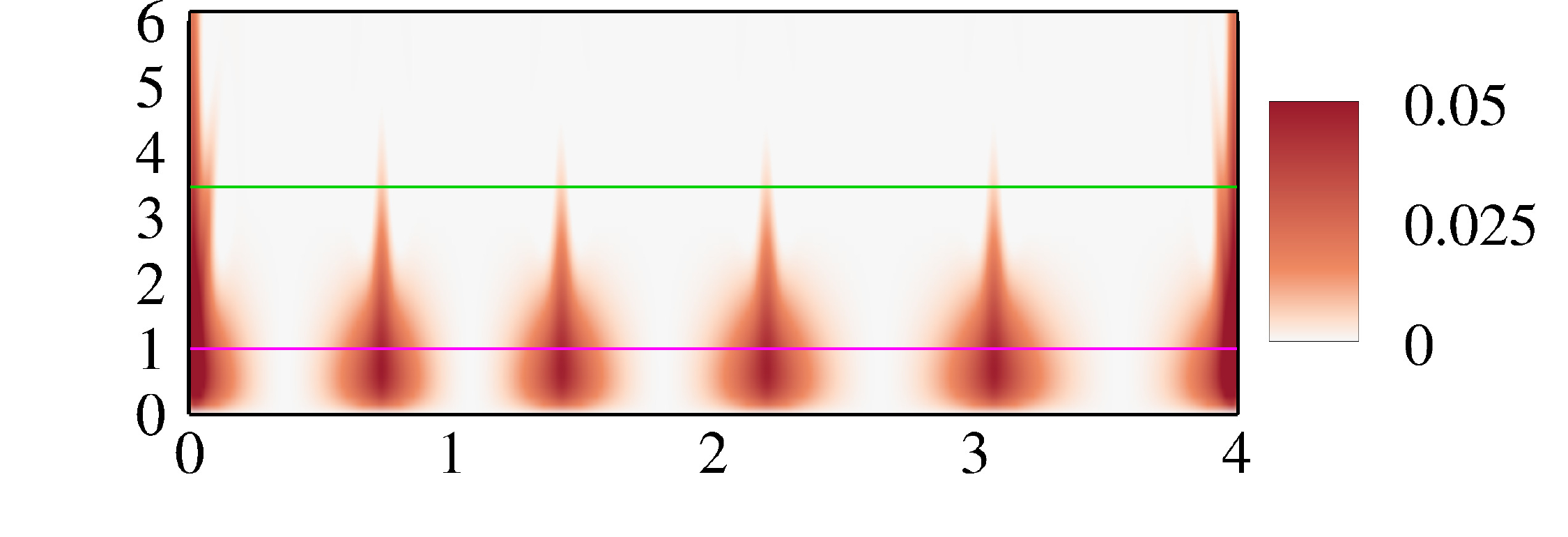}}
\put(3.0,0){$x$}
\put(0.14,1){\rotatebox{90}{$y/\delta_{\rm Sh}$}}
\put(0,2.1){\textit{c})}
\end{picture}
\hspace{0.1cm}
\begin{picture}(6.5,2.5)
\put(0,0){\includegraphics[trim={0.6cm 0.3cm 0cm 0cm},clip,width=5.65cm]{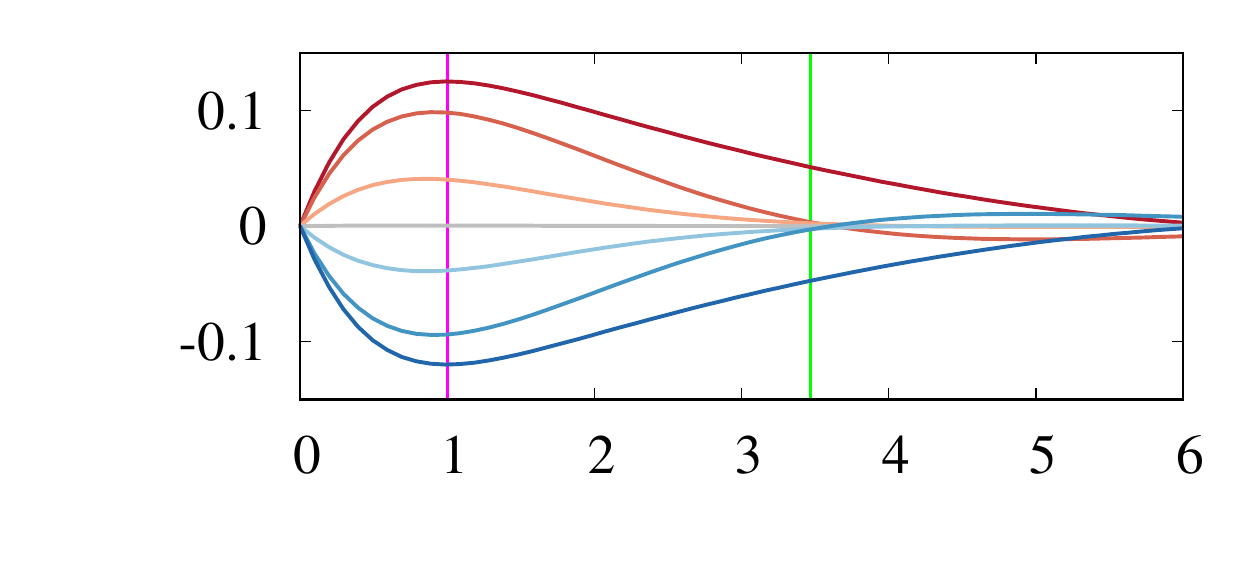}}
\put(3.0,0){$y/\delta_{\text{Sh}}$}
\put(0.25,1.1){\rotatebox{90}{$\langle u_z\rangle_{z}$}}
\put(0,2.1){\textit{d})}
\end{picture}
\vspace{0.1cm}
\caption{Detailed structure of the wall modes at $Ha=2000$. Horizontal cross section of the time-averaged (a) temperature 
field $T$, (b) vertical velocity component $u_z$ and (c) heat transport $u_zT$ at the mid-plane ($z=0$). The violet horizontal 
line indicates Shercliff boundary layer thickness $\delta_{\rm Sh}$. The green horizontal line indicates $r_{\rm cr}=3.47\delta_{\rm Sh}$. 
Sidewall distance $r_{\rm cr}$ is determined in figure \ref{crossover}. (d) Profiles of vertical velocity component versus the wall $y$-distance 
are taken at seven $x$-positions indicated by the cyan vertical lines in panel (b). Profiles from red to blue via grey correspond to the 
cyan lines from left to right. The vertical velocity component is averaged over the whole cell height in these profiles.}
\label{fig:ha2000}
\end{figure}
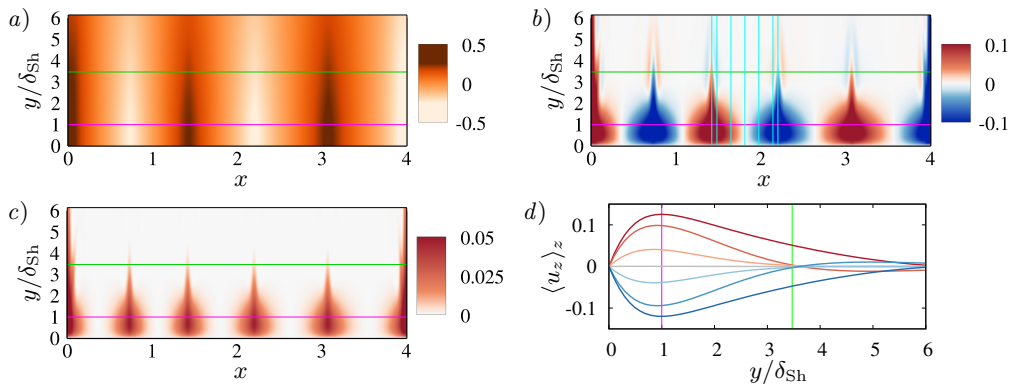
\begin{figure}
\centering
{\includegraphics[height=5.0cm]{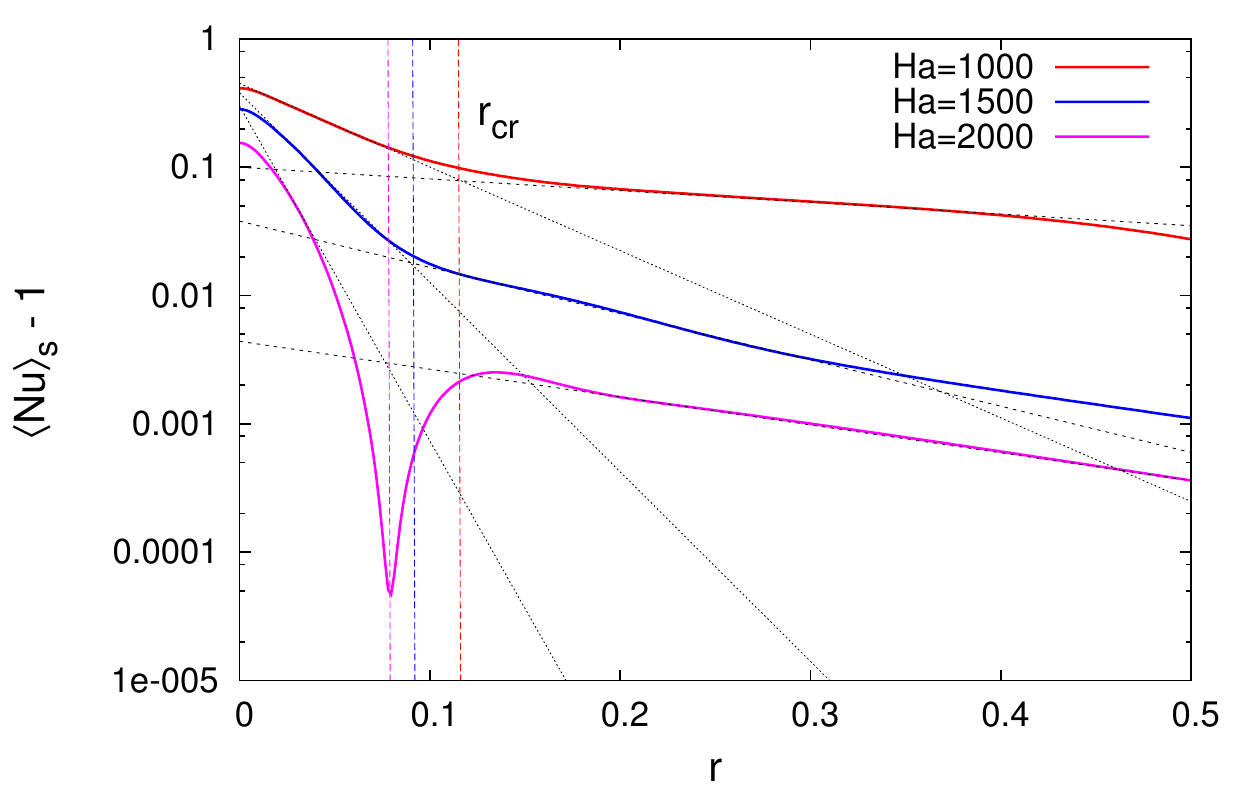}}
\caption{Two-layer structure of the wall modes. The global convective heat flux $Nu-1$ taken over cross sections $S$ is plotted with respect to 
the sidewall-normal coordinate $r$ is plotted for $Ha\ge 1000$. The spatial exponential decay with $r$ is fitted by two separate scaling laws 
which intersect at $r = r_{\rm cr}$.} 
\label{crossover}
\end{figure}

\section{Summary and discussion}
We have studied three-dimensional magnetoconvection in a closed rectangular cell under the influence of a strong external vertical magnetic field $B_0$. 
An increase of the magnetic field strength, which is measured by an increasing Hartmann number $Ha$, suppresses the highly turbulent motion of the enclosed 
liquid metal at $B_0=0$ ever stronger. In a close analogy to rotating RB convection, we find (laminar) sidewall modes that continue to exist for magnetic fields 
with $Ha>Ha_{\rm c}$. For the present set of simulations, we were able to follow these wall modes up to $Ha=2 Ha_{\rm c}$. A further increase of the Hartmann number
would require even finer mesh resolutions at the sidewalls. 
It is planned to ramp up these simulations for higher Hartmann numbers in the future.

A splitted jet or sandwich-type structure was seen in the linearly unstable modes of \cite{Houchens2002}. Our present simulations 
revealed this double-layer structure of the wall modes and show that it scales with the Shercliff layer thickness.  This again is similar to the boundary 
thicknesses near the sidewall in the rotating convection case where wall modes were found to be related to Stewartson layers that scale with Ekman number 
\citep{Kunnen2011,Kunnen2013}. We did not observe a drift of the wall modes which was observed in rotating RBC with different cell geometries
\citep{Knobloch1998,Vasil2008,Horn2017} and can be traced back to a breaking of azimuthal reflection symmetry \citep{Ecke1992}. It remains 
open which symmetry-breaking bifurcation could be at work for the convection flow in the presence of a strong magnetic field. The absence of a drift in our DNS might 
be attributed to the relatively short total integration time of 31 free fall time units which is a small fraction of the momentum diffusion time scale -- the slowest time in our flow on the basis of characteristic system parameters. At $Ra=10^7$ and $Pr=0.025$ this results to $t_{\rm vis}=\sqrt{Ra/Pr}\,T_f = 2\times 10^4 T_f$.  
Indeed, at Hartmann numbers of  $Ha\geq 1000$, magnetoconvection becomes a very slow dynamical process and numerical studies would require extremely long-term runs of this order of magnitude. 

Our present numerical findings for the existence of wall modes are consistent with the predictions by \cite{Houchens2002} and \cite{Busse2008} for the 
Hartmann number $\overline{Ha}_c$ at which the convection should be completely ceased in a closed cell. For $Ra=10^7$, this gives $\overline{Ha}_c=(Ra/68.25)^{2/3}\approx 2777$ if the asymptotic solution of \cite{Houchens2002} is taken from their closed cylindrical cell with $\Gamma=1$. From the 
asymptotic theory of \cite{Busse2008} that applies free-slip boundary conditions at the top and bottom plates follows $\overline{Ha}_c=(Ra/(3\pi^2\sqrt{3\pi/2}))^{2/3}\approx 2890$. Both theoretical approaches suggest thresholds that are still larger than the Hartmann number which could be obtained here. 
This should however be possible in a simulations which we plan to conduct in the near future, as already stated.       

The observed wall modes resemble also an interesting similarity to isolated turbulent spots in the Shercliff layers in MHD pipe 
and duct flows at the edge of relaminarization \cite[]{Krasnov:2013,ZikanovASME:2014}. Despite one major difference -- the present system 
is linearly unstable in contrast to pipe and square duct flows -- both cases lead to the development of residual structures that maintain a transport of heat and momentum.
They are formed in thin near-wall zones, whereas the rest of the domain remains essentially unperturbed.
Similar to MHD duct and pipe flows, wall modes are rather weak and, thus, difficult to identify in experiments if only integral parameters can be 
measured as discussed by \cite{Krasnov:2013}.  Figure~\ref{fig:z_profile}(b) and table \ref{tab:kd} show that these modes provide virtually
no impact on the vertical temperature distribution and the Nusselt number only slightly differs from the lower diffusive bound. These similarities 
suggest that the residual sidewall structures are very likely a common feature of MHD wall bounded flows subject to strong external magnetic fields.

\vspace{0.3cm}
WL is supported by the Deutsche Forschungsgemeinschaft with Grant No. GRK 1567 and by a Fellowship of the China Scholarship Council.
DK acknowledges support by the Deutsche Forschungsgemeinschaft with Grant SCHU 1410/29. 
Computer time has been provided by Large Scale Project pr62se of the Gauss Centre for Supercomputing at the SuperMUC cluster at the 
Leibniz Rechenzentrum Garching. We thank Till Z\"urner,  Andr\'{e} Thess and Christian Karcher for discussions and suggestions. The work 
of JS is also supported by the Tandon School of Engineering at New York University.

\bibliographystyle{jfm}
\bibliography{reference}

\end{document}